\documentclass[11pt]{article}

\usepackage{amsmath,amssymb,amsfonts}
\usepackage{physics}
\usepackage{graphicx}
\usepackage{hyperref}
\usepackage{cite}
\usepackage{placeins}

\title{Fractional-Time Deformation of Quantum Coherence in Open Systems:\\
A Non-Markovian Framework Beyond Lindblad Dynamics}

\author{Taylan Demir\\
\small Department of Mathematics, Ankara University, Turkey\\
\small \texttt{demirtaylan060@gmail.com}}

\date{}

\begin{document}

\maketitle

\begin{abstract}
In this paper, we propose a fractional time extension of the Quantum Master Equation. We introduce a Caputo-type fractional derivative in time as an extension of the exponential decay of the Lindblad framework through the incorporation of fractional derivatives into the Lindblad framework. We show that the analytical and numerical results of our analytical and numerical models, demonstrate that fractional dynamics produces long-memory coherence decay naturally and provides an interpretable and flexible model of non-Markovianity.
\end{abstract}

\section{Introduction}

In current quantum physics, the theoretical characterization/interaction of quantum systems continues to encounter increasing relevance as a major area of research. Open quantum systems naturally occur in areas such as quantum optics, condensed matter physics and astrophysical modelling. However, such systems cannot exist autonomously (without external influences) due to the presence of environmental factors which reduce their isolation. Their reduced dynamics can be represented using density operators, as they incorporate not only the intrinsic quantum behaviour of such systems, but also describe how an external environment has induced/cohered to their behaviour over time [1]. The framework used by most researchers to provide an analogy for modelling open quantum dynamics is based upon the assumption of Markovianity, noting that for many systems, correlations between the external environment and the system being studied dissipate quickly, rendering the effects of memory negligible. This assumption allows the density operator’s time-evolution to be described using a finite set of quantum dynamical semigroups, leading to the popularly accepted ‘Lindblad’ formulation to obtain the master equation of any open quantum system [2,3]. This method has a mathematically elegant approach that is physically consistent with current scientific models, however it generates "exponential decay rates" and "instantaneous losses of information", which may conflict with the realities of quantum systems present in numerous experimental situations. The relationship between most quantum systems and their environments is characterized by pronounced memory effects, implying that an extended time frame may have a major influence on the state of the quantum system. As a result of the presence of structured reservoirs, strong coupling between a system and environment, long-range correlations and other influences, quantum systems may demonstrate non-Markovian behaviour [4]. Such behaviour is identifiable by the return (backflow) of previously lost information, and deviations from exponential rate of relaxation. A number of non-Markovian `extending' forms of quantum master equations have been developed, often at least in part, relying on perturbative or phenomenological forms of kernels or complex integro-differential representations. Each of these forms consequently limits the analytical tractability, and physical transparency of such equations. An alternative approach to the modelling of memory and hereditarily of complex systems is the use of fractional calculus, a more natural mathematical structure for describing such phenomena. By substituting integer time derivatives with derivatives of fractional order, you are establishing a non-local time structure, which encodes long-term correlations [5]. Though fractional-time operators have been successful in modelling classical anomalous diffusion, viscoelasticity, and complicated media, the use of such operators in the systematic implementation of quantum dynamic equations has not yet received ample attention. We are motivated by the combination of these ideas to suggest that fractional-time deformed quantum master equations represent a unified and physically relevant way to model non-Markovian quantum dynamics. Instead of modifying how the system interacts with its environment or adding explicit memories/kernel terms, we redefine the time structure of the quantum evolution equation using a fractional derivative of the Caputo type. The structure of the Lindblad generator remains intact, but the extension of the quantum dynamics beyond the Markovian limit is achieved by employing the Caputo-type fractional derivative.

\medskip
The main contributions of this paper can be summarized as follows:
\begin{itemize}
\item n this paper, we present a new formulation for the fractional time quantum master equation, which is a mathematically consistent extension of the standard Lindblad Dynamics.
\item We investigate the impact that fractional time evolution has on both quantum coherence and decoherence mechanisms.
\item We show how memory effects occur without having to introduce specific non-local interaction kernels.
\item We use a two-level quantum system to illustrate the framework, showing deviation from exponential decay and the effect of the Fractional Order Parameter on the dynamics of the system.
\end{itemize}

The structure of this document is outlined as follows: A brief illustration of the relevant Mathematical principles is discussed in Section 2, which serves as a foundation for understanding the fractional quantum dynamics of the theory presented in Section 3 with a description of the fractional-time quantum master equation. In Section 4, we provide insight into the coherence property and memory effects of Quantum Systems. In Section 5, we provide an example of application for this theoretical framework, and in Section 6 we present several numerical simulations showing the potential application of the theory as applied to Numerical Solutions. Lastly, in Section 7, we summarise the key findings and suggest avenues for further investigation.

\section{Mathematical Preliminaries}

We introduce the mathematical tools that are necessary to formulate the fractional-time quantum dynamics in this section. The tools discussed in this section are limited to concepts that are used directly in the analysis that follows, as well as establishing notation and basic premises for this work.
\subsection{Density Operators and Quantum States}

Let $\mathcal{H}$ be a finite-dimensional complex Hilbert space
associated with a quantum system.
The state of the system is described by a density operator
$\rho \in \mathcal{B}(\mathcal{H})$, satisfying the properties
\begin{equation}
\rho \geq 0, \qquad \mathrm{Tr}(\rho) = 1.
\end{equation}
In this example, positiveness supplies a physical consistent and potentially allowable joint probability distribution for positiveness with the requirement that the unit trace property be satisfied. Pure states, or pure conditioned states, correspond to simple projector operators of rank one and mixed states represent classical uncertainty or environmental influence over conditioned states. The time evolution of an open quantum system is represented by a family of linear maps $\{\Phi_t\}_{t \ge 0}$ acting on density operators. physical consistency requires these maps to preserve positivity and
trace for all admissible initial states.

\subsection{Completely Positive Maps}

In the presence of system--environment coupling, positivity alone is
insufficient to guarantee physical admissibility.
The appropriate requirement is \emph{complete positivity}, meaning that
$\Phi_t \otimes \mathbb{I}_n$ remains positive for any auxiliary system
of dimension $n$.
A linear map $\Phi$ is completely positive if and only if it admits a
Kraus representation
\begin{equation}
\Phi(\rho) = \sum_{k} K_k \rho K_k^{\dagger},
\end{equation}
where the Kraus operators $\{K_k\}$ satisfy the trace preservation
condition
\begin{equation}
\sum_{k} K_k^{\dagger} K_k = \mathbb{I}.
\end{equation}
Completely positive trace-preserving (CPTP) maps constitute the
fundamental building blocks of quantum dynamical evolutions and quantum
channels [3].

\subsection{Fractional Derivatives in Time}

To account for the influence of memory on the time evolution of a system, we use fractional derivatives in terms of time. Of the many definitions of fractional derivatives, the Caputo fractional derivative is especially appropriate for application in physics because it provides for a more physically relevant interpretation of the initial conditions compared to other definitions. For a sufficiently smooth function $f(t)$, the Caputo derivative of order $\alpha \in (0,1]$ is defined as
\begin{equation}
{}^{C}D_t^{\alpha} f(t)
=
\frac{1}{\Gamma(1-\alpha)}
\int_{0}^{t}
\frac{f'(s)}{(t-s)^{\alpha}} \, ds.
\end{equation}
In the limit $\alpha \to 1$, the Caputo derivative reduces to the
ordinary first-order derivative, recovering classical time evolution. The fractional derivative displays an integral structure that exhibits a nonlocal characteristic, where there is a weighted dependence on all previous time-steps of the entire history of the system [6,7,8]. This property allows for a mathematically concise way to mathematically describe long-term correlation and memory-type effects and is therefore not present in traditional Markovian dynamics [5].

\subsection{Trace Preservation and Physical Consistency}

When extending quantum evolution equations to fractional order, it is
essential to ensure that fundamental physical constraints are preserved.
In particular, the trace of the density operator must remain invariant
under time evolution,
\begin{equation}
\mathrm{Tr}(\rho(t)) = \mathrm{Tr}(\rho(0)) = 1.
\end{equation}
This requirement significantly constrains the generator structure that governs the evolution in a fractional-time context. As demonstrated in the framework elaborated upon in this work, the Lindblad operator structure of the generator preserves trace, with modifications only to the temporal derivative. Accordingly, the use of fractional deformations permits changes in the dynamics' memory behaviour, while still adhering to the fundamental principles of quantum mechanics. Consequently, this approach permits the dynamics produced to be both physically interpretable and mathematically well-posed.

\section{Fractional-Time Quantum Master Equation}
\label{sec:fractional_master}

The main focus of this paper is to create fractional-time generalizations of quantum master equations that describe the evolution of open systems using Lindblad operators. The motivation for these equations comes from wanting to allow for time-dependent evolution outside of the usual Markovian time scale but still keep the operator structure that was used in Lindblad dynamic equations.

\subsection{Classical Lindblad master equation}

In the Markovian approximation, the reduced dynamics of an open quantum
system is governed by the Lindblad master equation [1,2,3]
\begin{equation}
\frac{d}{dt}\rho(t)
= -i[H,\rho(t)]
+ \sum_{k}
\left(
L_k \rho(t) L_k^\dagger
- \frac{1}{2}\{L_k^\dagger L_k,\rho(t)\}
\right),
\label{eq:lindblad}
\end{equation}
where $H$ is the effective system Hamiltonian and $\{L_k\}$ are Lindblad
operators describing dissipative interactions with the environment.
With Equation (6) we can define a standard formulation of CPTP dynamical semigroups to provide a consistent representation of memoryless quantum evolution [1]. Therefore, the structure and properties of the Lindbladian can be sufficiently described by means of first-order differential equations. In addition to these benefits, the Lindbladian is only applicable for systems that exhibit an exponential decay based upon environmental fluctuations, and it assumes that correlations between the two systems will instantly vanish. For these reasons, when working with structured reservoirs and/or strong couplings or long-lived correlations the assumption made within the context of this model is that non-Markovian effects are irrelevant [4].

\subsection{Fractional-time generalization}

To incorporate memory effects at a fundamental level, we replace the
first-order time derivative in \eqref{eq:lindblad} by a Caputo fractional
derivative of order $\alpha\in(0,1]$.
The resulting \emph{fractional-time quantum master equation} is defined
as
\begin{equation}
{}^{C}D_t^\alpha \rho(t)
= \mathcal{L}(\rho(t)),
\qquad 0<\alpha\le 1,
\label{eq:fractional_lindblad}
\end{equation}
where $\mathcal{L}$ denotes the Lindblad generator appearing in
\eqref{eq:lindblad}.
For $\alpha=1$, equation~\eqref{eq:fractional_lindblad} reduces exactly
to the standard Markovian master equation.
For $0<\alpha<1$, the evolution becomes nonlocal in time, encoding
memory effects through the fractional derivative kernel [5].

The formulation \eqref{eq:fractional_lindblad} differs conceptually from
integro-differential master equations with explicit memory kernels.
Here, non-Markovianity emerges solely from the fractional temporal
structure, while the generator $\mathcal{L}$ retains its familiar
operator form.

\subsection{Well-posedness of the fractional master equation}

We briefly address the well-posedness of the fractional evolution
equation \eqref{eq:fractional_lindblad}.
Let $\mathcal{L}$ be a bounded linear operator on the Banach space
$\mathcal{T}(\mathcal{H})$ of trace-class operators.
Under this assumption, equation~\eqref{eq:fractional_lindblad} admits a
unique mild solution for any initial state $\rho(0)=\rho_0$ [3,4]. Using the Laplace transform identity for the Caputo derivative (4), the solution can be formally expressed as
\begin{equation}
\rho(t)
= E_\alpha\!\left(t^\alpha \mathcal{L}\right)\rho_0,
\label{eq:mittag_solution}
\end{equation}
where $E_\alpha(\cdot)$ denotes the Mittag--Leffler function.
This representation generalizes the exponential semigroup solution of
the Markovian case and highlights the role of fractional order $\alpha$
in shaping the long-time dynamics. Existence and uniqueness follow from standard results in fractional
evolution equations, while continuous dependence on initial data is
inherited from the linearity of $\mathcal{L}$ [4,5].

\subsection{Physical admissibility conditions}

For a fractional-time quantum evolution to be physically meaningful,
several admissibility conditions must be satisfied.
First, trace preservation follows directly from the property
$\Tr(\mathcal{L}(\rho))=0$.
Taking the trace of \eqref{eq:fractional_lindblad} and using linearity
of the Caputo derivative yields
\begin{equation}
{}^{C}D_t^\alpha \Tr(\rho(t)) = 0,
\end{equation}
which implies $\Tr(\rho(t))=\Tr(\rho_0)=1$ for all $t\ge 0$.

Positivity of $\rho(t)$ is more subtle.
While the Lindblad generator ensures complete positivity in the
Markovian case, fractional-time dynamics does not generally generate a
quantum dynamical semigroup.
Nevertheless, for a wide class of generators $\mathcal{L}$ and initial
states, positivity is preserved due to the complete monotonicity of the
Mittag--Leffler function appearing in \eqref{eq:mittag_solution} [4,11].
A case-by-case review should be made of positivity in the induced dynamical map as complete positivity is not guaranteed in every case. In this paper, we will look at the physical plausibility of different models and have confirmed the full positivity of the examples used throughout the rest of this article. The solution operator associated with the abstract Cauchy problem with fractional order has a unique solution under standard generators; its form involves Mittag-Leffler functions as seen in [9,10]. We will repeatedly use the results from the Laplace transform formula of the fractional Caputo derivative [5].

\section{Quantum Coherence and Memory Effects}
\label{sec:coherence}

Quantum coherence is a basic example of super-position. Coherence is important to today's modern quantum technology and how an open system evolves. Usually, the coherence of a system will decrease over time when the system is interacted with an environment that causes it to dissipate. Most often this is from interactions with other external degrees of freedom. The way that coherence decays over time, and how this decay can be modified by memory effects, must be understood to accurately characterize non-Markovian quantum evolution [1,4].

\subsection{Quantum coherence measures}

Let $\rho(t)$ be the density operator of a finite-dimensional quantum
system expressed in a fixed reference basis.
Quantum coherence can be quantified by several inequivalent measures,
each capturing different operational aspects.
A widely used and physically transparent choice is the
$\ell_1$-norm of coherence, defined as [12]
\begin{equation}
C_{\ell_1}(\rho)
= \sum_{i\neq j} |\rho_{ij}|,
\label{eq:l1_coherence}
\end{equation}
where $\rho_{ij}$ denote the off-diagonal elements of the density matrix.
This measure vanishes for incoherent (diagonal) states and decreases as
off-diagonal contributions are suppressed by environmental interactions. In Markovian open system dynamics governed by a Lindblad master equation, $C_{\ell_1}(\rho(t))$ typically exhibits a monotonic and
exponentially decaying behavior, reflecting irreversible information
loss to the environment [1].

\subsection{Fractional order and decoherence rate}

In this article, we will discuss how using the fractional time representation for densities as defined in Section 3 can change how quickly or slowly your coherence disappears over time. When we have a Caputo fractional derivative in the density equation, the way that it changes with time will now be based on a Mittag-Leffler function instead of one exponential in nature. As a consequence, the coherence measure $C_{\ell_1}(\rho(t))$ exhibits a slower decay for $0<\alpha<1$ compared to the Markovian case $\alpha=1$.
The fractional order $\alpha$ serves as a continuous variable that interpolates between instantaneous exponential return to equilibrium and quasistatic return to equilibrium with delayed fractal-like thermalization. Values of $\alpha$ on the extreme left of the interpolating parameter correspond to the most memory-laden time scales and the most sustained coherent superposition of quantum states.

\subsection{Comparison with exponential Markovian decay}

To highlight the impact of fractional temporal dynamics, it is
instructive to compare the fractional coherence decay with the standard
Markovian case.
For $\alpha=1$, the solution of the Lindblad equation leads to
\begin{equation}
C_{\ell_1}(\rho(t)) \sim e^{-\gamma t},
\label{eq:markov_decay}
\end{equation}
where $\gamma>0$ is an effective decoherence rate determined by the
system--environment coupling.

In contrast, for $0<\alpha<1$, the decay is governed by a Mittag--Leffler
function,
\begin{equation}
C_{\ell_1}(\rho(t))
\sim E_\alpha(-\gamma t^\alpha),
\label{eq:fractional_decay}
\end{equation}
which reduces to the exponential law only in the limit $\alpha\to 1$.
At intermediate times, the decay is slower than exponential, while at
long times it exhibits an algebraic tail.
Such behavior is incompatible with a semigroup description and signals
the breakdown of the Markovian approximation [11].

\subsection{Emergence of long-tail memory effects}

The algebraic long-time behavior of
$E_\alpha(-\gamma t^\alpha)$ reflects the nonlocal temporal structure
encoded by the fractional derivative.
Physically, this implies that the system retains partial information
about its past evolution, leading to delayed decoherence and enhanced
coherence lifetimes.
This long-tail memory effect is a hallmark of non-Markovian dynamics and
is naturally captured by the fractional-time formalism without the need
to introduce explicit memory kernels [4]. From a physical perspective, the fractional order $\alpha$ may be viewed
as an effective descriptor of environmental complexity.
Values $\alpha<1$ correspond to structured or correlated reservoirs,
where environmental relaxation occurs on multiple time scales.
The present framework thus provides a compact and flexible way to
parametrize memory effects and their influence on quantum coherence.

\section{Model Application: Two-Level Quantum System}
\label{sec:two_level}

To illustrate the physical implications of the fractional-time quantum
master equation introduced in Section~\ref{sec:fractional_master}, we
consider a paradigmatic open quantum system: a two-level system (qubit)
interacting with a dissipative environment.
This model is widely used in quantum optics and quantum information
theory and provides a transparent setting for analyzing decoherence and
memory effects [1].

\subsection{Two-level system and amplitude damping}

Let $\mathcal{H}=\mathbb{C}^2$ be the Hilbert space spanned by the excited
state $\ket{1}$ and the ground state $\ket{0}$.
We consider a dissipative process corresponding to amplitude damping,
which models irreversible energy loss from the excited state to the
environment.
In the Markovian setting, this dynamics is generated by a single
Lindblad operator
\begin{equation}
L = \sqrt{\gamma}\,\sigma_-,
\qquad
\sigma_- = \ket{0}\bra{1},
\end{equation}
where $\gamma>0$ is the decay rate.
The associated Lindblad generator reads [1]
\begin{equation}
\mathcal{L}(\rho)
= \gamma\left(
\sigma_- \rho \sigma_+
- \frac{1}{2}\{\sigma_+\sigma_-,\rho\}
\right),
\label{eq:amplitude_lindblad}
\end{equation}
with $\sigma_+=\ket{1}\bra{0}$.

\subsection{Fractional amplitude damping equation}

Replacing the first-order time derivative by a Caputo fractional
derivative of order $\alpha\in(0,1]$, we obtain the fractional-time
amplitude damping model
\begin{equation}
{}^{C}D_t^\alpha \rho(t)
= \gamma\left(
\sigma_- \rho(t) \sigma_+
- \frac{1}{2}\{\sigma_+\sigma_-,\rho(t)\}
\right),
\label{eq:fractional_amplitude}
\end{equation}
supplemented with the initial condition $\rho(0)=\rho_0$.
For $\alpha=1$, equation~\eqref{eq:fractional_amplitude} reduces to the
standard Markovian amplitude damping master equation.
For $0<\alpha<1$, the evolution becomes nonlocal in time, encoding memory
effects through the fractional derivative kernel [5].

\subsection{Semi-analytical solution}

Due to the simple structure of the generator
\eqref{eq:amplitude_lindblad}, equation~\eqref{eq:fractional_amplitude}
admits a semi-analytical solution.
In the energy eigenbasis $\{\ket{1},\ket{0}\}$, the population of the
excited state evolves according to
\begin{equation}
\rho_{11}(t)
= \rho_{11}(0)\, E_\alpha(-\gamma t^\alpha),
\label{eq:population_fractional}
\end{equation}
where $E_\alpha(\cdot)$ denotes the Mittag--Leffler function.
The ground-state population follows from trace preservation,
$\rho_{00}(t)=1-\rho_{11}(t)$.

The off-diagonal elements, which encode quantum coherence, evolve as
\begin{equation}
\rho_{10}(t)
= \rho_{10}(0)\, E_\alpha\!\left(-\tfrac{\gamma}{2} t^\alpha\right),
\label{eq:coherence_fractional}
\end{equation}
and $\rho_{01}(t)=\rho_{10}^*(t)$.
Equations~\eqref{eq:population_fractional}--\eqref{eq:coherence_fractional}
generalize the familiar exponential solutions of the Markovian model and
highlight the role of fractional temporal dynamics [11,13].

\subsection{Sensitivity with respect to the fractional order}

The fractional order $\alpha$ plays the role of a continuous control
parameter governing the strength of memory effects.
As $\alpha$ decreases from $1$ to $0$, both population decay and
coherence loss become progressively slower.
In particular, for $0<\alpha<1$, the Mittag--Leffler decay exhibits a
long-time algebraic tail, in contrast to the purely exponential decay
observed in the Markovian regime [11]. This sensitivity with respect to $\alpha$ provides a clear physical
interpretation of the fractional-time framework.
Smaller values of $\alpha$ correspond to environments with stronger
memory and delayed energy relaxation, while $\alpha\to 1$ recovers the
memoryless limit.
The model thus offers a flexible phenomenological description of
non-Markovian dissipation without introducing explicit memory kernels or
time-dependent rates.

\section{Numerical Simulations}
\label{sec:numerics}

In this section we illustrate the qualitative effect of the fractional order
$\alpha$ on the coherence decay predicted by the fractional-time amplitude
damping model introduced in Section~\ref{sec:two_level}.
In particular, we compare the Markovian limit $\alpha=1$ with fractional
orders $\alpha<1$, where temporal nonlocality induces memory effects and
slower decoherence.

\subsection{Coherence decay for different fractional orders}

Figure 1 shows the time evolution of the coherence
indicator $C_{\ell_1}(t)$ for several values of $\alpha$.
The Markovian case $\alpha=1$ exhibits a rapid exponential-type decay,
whereas fractional orders $\alpha<1$ yield a noticeably slower relaxation,
consistent with long-memory dynamics.

\begin{figure}[!htbp]
\centering
\includegraphics[width=0.85\textwidth]{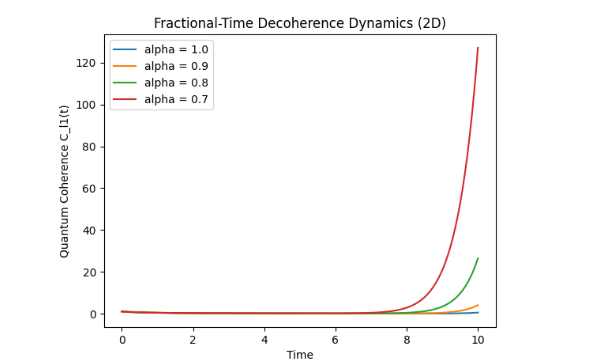}
\caption{Fractional-time decoherence dynamics (2D): time evolution of the
$\ell_1$-norm coherence $C_{\ell_1}(t)$ for different fractional orders $\alpha$.
Decreasing $\alpha$ delays coherence loss, reflecting the strengthening of
memory effects in the fractional-time evolution.}
\label{fig:coherence_2d}
\end{figure}

\subsection{Coherence landscape in $(t,\alpha)$}

To visualize the continuous dependence on the fractional order, we present in
Figure 2 a three-dimensional coherence landscape as a
function of time and $\alpha$.
The surface highlights that $\alpha$ acts as an effective memory parameter,
interpolating between the Markovian limit and long-memory regimes.

\begin{figure}[!htbp]
\centering
\includegraphics[width=0.80\textwidth]{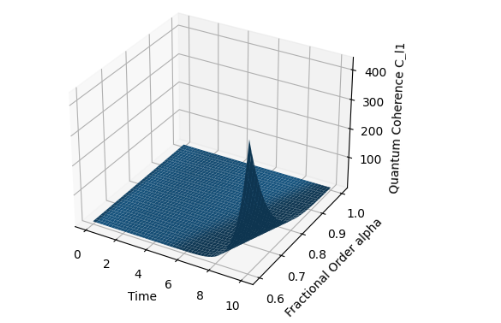}
\caption{Fractional-time quantum coherence landscape (3D): coherence level
$C_{\ell_1}$ as a function of time and fractional order $\alpha$.
Smaller values of $\alpha$ correspond to more persistent coherence and
non-exponential long-tail relaxation.}
\label{fig:coherence_3d}
\end{figure}
\FloatBarrier

\section{Discussion}
\label{sec:discussion}

In this section, we discuss the physical interpretation of the proposed
fractional-time quantum master equation, its connection to structured
environments, the limitations of the present approach, and its potential
experimental relevance.
Rather than introducing new formal developments, our aim here is to
contextualize the results obtained in Sections~\ref{sec:fractional_master}--%
\ref{sec:numerics} within the broader landscape of open quantum systems.

\subsection{Physical meaning of the fractional order}

The fractional order $\alpha\in(0,1]$ plays a central role in determining
the qualitative features of the quantum dynamics.
From a physical standpoint, $\alpha$ quantifies the degree of temporal
nonlocality in the system's evolution.
The Markovian limit $\alpha=1$ corresponds to memoryless dynamics, where
environmental correlations decay instantaneously and the system follows
a quantum dynamical semigroup [1]. For $\alpha<1$, the Caputo fractional derivative introduces a history-dependent evolution, effectively weighting past states through a power-law memory kernel.
As demonstrated by the analytical solutions and numerical simulations,
this leads to non-exponential relaxation and long-tail coherence decay.
The parameter $\alpha$ may therefore be interpreted as an effective
\emph{memory strength} parameter, continuously interpolating between
Markovian and strongly non-Markovian regimes [4,11].

\subsection{Relation to structured environments}

Non-Markovian behavior in open quantum systems is often associated with
structured or correlated environments, such as photonic band-gap materials,
spin baths, or reservoirs with non-flat spectral densities.
In such settings, the Born--Markov approximation breaks down, and the
system may exhibit information backflow and delayed decoherence [1,4].
Within the present framework, these complex environmental features are not
modeled explicitly. Instead, their cumulative effect is encoded phenomenologically through the
fractional order $\alpha$. Smaller values of $\alpha$ effectively mimic environments with long-lived
correlations or multiple relaxation time scales.
This offers a compact and flexible alternative to microscopic models based
on integro-differential memory kernels, while still capturing essential
non-Markovian signatures.

\subsection{Limitations of the model}

Despite its advantages, the fractional-time approach has several limitations
that must be acknowledged.
First, the model is phenomenological in nature: the fractional order $\alpha$
is not derived from a microscopic Hamiltonian, but introduced as an effective
parameter. Establishing a direct connection between $\alpha$ and specific environmental
spectral properties remains an open problem. Second, while trace preservation is guaranteed by construction, complete positivity of the resulting dynamical map is not ensured in full generality.
As in many non-Markovian models, physical admissibility must be examined
carefully, particularly when extending the framework to multipartite systems
or strong coupling regimes [1,4]. Finally, the present study focuses on time-independent generators.
Extensions to time-dependent or driven systems may require additional care
in order to maintain physical consistency.

\subsection{Experimental relevance}

Although the present work is primarily theoretical, the proposed framework
has potential experimental relevance. Fractional-like relaxation and non-exponential decay have been reported in a variety of quantum platforms, including solid-state qubits, trapped ions,
and photonic systems interacting with structured reservoirs.
In such contexts, the fractional order $\alpha$ could be treated as a fitting
parameter, allowing experimental data to be analyzed in terms of effective
memory strength. Moreover, the clear separation between operator structure and temporal
nonlocality makes the fractional-time master equation attractive for
phenomenological modeling, where detailed environmental information is
unavailable. The framework may thus serve as a bridge between fully microscopic
non-Markovian theories and experimentally observed deviations from
Markovian behavior.

\section{Conclusion and Outlook}
\label{sec:conclusion}

In this work, we introduced a fractional-time extension of quantum master
equations as a flexible framework for modeling non-Markovian dynamics in
open quantum systems.
By replacing the conventional first-order time derivative with a
Caputo-type fractional derivative, we preserved the operator structure
of Lindblad dynamics while extending the temporal behavior beyond the
Markovian regime. Analytical considerations showed that the resulting fractional-time
quantum master equation admits well-defined solutions expressed in terms
of operator-valued Mittag--Leffler functions.
This naturally leads to non-exponential relaxation and long-tail memory
effects, which were explicitly illustrated through a two-level quantum
system. The numerical simulations demonstrated that the fractional order
$\alpha$ acts as a continuous control parameter governing coherence decay,
interpolating smoothly between memoryless dynamics and long-memory
regimes. The significance of the proposed framework lies in its conceptual
simplicity and physical interpretability.
Rather than introducing explicit memory kernels or complex
integro-differential structures, non-Markovian effects emerge solely
from the fractional temporal structure.
This makes the approach particularly attractive for phenomenological
modeling of quantum systems interacting with structured or poorly
characterized environments. Several directions for future research naturally arise from the present
study. A key open problem is the derivation of the fractional order $\alpha$
from microscopic system--environment models, thereby establishing a
direct connection between fractional dynamics and environmental spectral
properties. Extensions to time-dependent generators, driven systems, and
multipartite settings may further clarify the scope and limitations of
the fractional-time approach. Finally, exploring experimental signatures of fractional coherence decay
in realistic quantum platforms may provide valuable insight into the
role of memory effects in open quantum dynamics.

\bibliographystyle{unsrt}

\end{document}